\documentstyle[12pt,epsf]{article}

\let\To\rightarrow
\newcommand{\epem} {e^+e^-}
\newcommand{\EPR}  {\epsilon^{\prime}/\epsilon}
\newcommand{\KL}  {K_{\rm L}}
\newcommand{\KS}  {K_{\rm S}}
\newcommand{\KZ} {K^0}
\newcommand{\KZB} {{\overline K}^0}
\newcommand{\ptsq}{\mbox{$p_t^2$}}

\def\etal{{\it et al.}}
\def\ie{{\it i.e.}}

\def\Title#1{\begin{center} {\Large #1 } \end{center}}

\begin{document}

\Title{$\EPR$ Results from KTeV}

\bigskip\bigskip


\begin{raggedright}  

{\it E. Blucher\index{Blucher, E.}\\
The Enrico Fermi Institute \\
The University of Chicago\\
Chicago, IL 60637 USA}
\bigskip\bigskip
\end{raggedright}

\section{Introduction}

Since the 1964 discovery of 
$\KL\To\pi^+\pi^-$~\cite{Christenson:1964fg},
understanding
CP violation has been a major goal of particle physics.
Subsequent experiments showed that
the dominant
mechanism of CP violation in neutral kaons (still the only particle system
in which CP violation has been observed) is a small asymmetry between
$\KZ \To \KZB$ and $\KZB \To \KZ$ mixing.  This asymmetry, 
referred to as indirect CP violation,
results 
in the $K_L$ and $K_S$ being states of mixed CP.
The 
parameter $\epsilon$, which is used to parametrize this effect, 
quantifies the CP impurity of the $K_L$ and $K_S$ states:
\begin{eqnarray*}
K_S &\sim & K_{\rm even} + \epsilon K_{\rm odd} \\
K_L &\sim & K_{\rm odd} + \epsilon K_{even},
\end{eqnarray*}
where $|\epsilon| = 2.28 \times 10^{-3}$,
$CP|K_{\rm even}> = +1|K_{\rm even}>$,
and $CP|K_{\rm odd}> = -1|K_{\rm odd}>.$

There has been considerable effort during the last 30 years to determine
whether or not the CP symmetry is also violated in the decay amplitude 
(\ie, $K_{\rm odd} \To \pi\pi$). This effect is referred to as direct CP
violation and is parametrized by $\epsilon^{\prime}$.
The ratio $\epsilon'/\epsilon$ 
can be determined from the double ratio of the
2-pion decay rates of $K_L$ and $K_S$:
\begin{equation}
{\rm Re}(\epsilon'/\epsilon) \simeq \frac{1}{6}\left[ \frac
{\Gamma(K_L \To \pi^+\pi^-)/\Gamma(K_S \To \pi^+\pi^-)}
{\Gamma(K_L \To \pi^0\pi^0)/\Gamma(K_S \To \pi^0\pi^0)}-1\right] .
\end{equation}
$\EPR \ne 0$ is an unambiguous indication of direct CP violation.

The Standard Model predicts both direct and indirect CP violation.  
Unfortunately, large hadronic uncertainties make precise calculations of
Re$(\EPR)$ difficult. 
Most recent Standard Model predictions~\cite{epstheory} are in the range
Re$(\EPR) = (0-30) \times 10^{-4}$.
Other models,
such as the Superweak Model of Wolfenstein~\cite{Wolfenstein:1964ks},
predict no direct CP violating effects. 

The two best previous measurements of $\EPR$ come from 
E731~\cite{Gibbons:1993zq} at Fermilab 
and NA31~\cite{Barr:1993rx} at CERN:
\begin{eqnarray*}
{\rm Re}(\epsilon'/\epsilon) & = & (7.4 \pm 5.9) \times 10^{-4} \ ({\rm E731}) \\
{\rm Re}(\epsilon'/\epsilon) & = &(23 \pm 6.5) \times 10^{-4}  \  ({\rm NA31}).
\end{eqnarray*}
The CERN result is 3.5 standard deviations
from zero, while the Fermilab result is just over
1 sigma from zero. 
To clarify this experimental situation and definitively resolve the
question of whether or not direct CP violation occurs,
new experiments were built at Fermilab (KTeV), CERN (NA48), and 
Frascati (KLOE) to attempt to 
measure $\EPR$ at the $(1-2) \times 10^{-4}$
level.  
The KTeV and NA48 experiments are similar fixed-target 
experiments. They differ mainly in the method used to produce $\KS$, 
and in the 
technique used to correct for the difference in detector acceptance for $\KS$ 
and $\KL$ decays
resulting from
the large $\KS-\KL$ lifetime difference. 
The KLOE experiment
at Frascati is trying a new technique using an $\epem \To \phi$
collider. 

The first measurement of $\EPR$ from the KTeV experiment~\cite{ktevcollab} 
is  the subject of this paper. A more complete description of the KTeV
measurement is given in~\cite{Alavi-Harati:1999xp}. The NA48 and 
KLOE experiments are described by Barr~\cite{barr} and 
Bertolucci~\cite{bertolucci} in these proceedings.

\section{The KTeV Detector}

\begin{figure}	
\centerline{\epsfxsize 5 truein 
\epsfbox{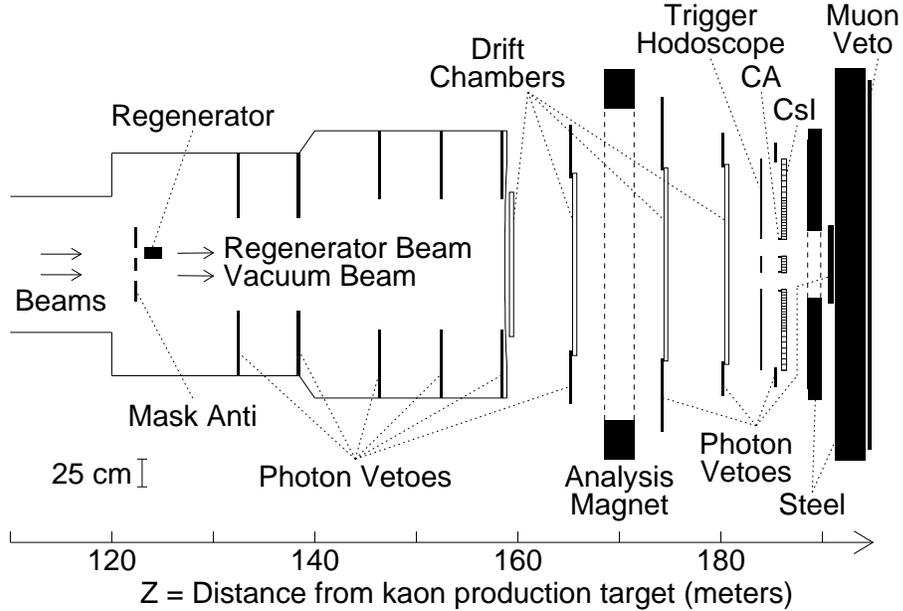}}
\vskip -.7 cm
\caption{Diagram of the KTeV detector.  
\label{fi:ktev}}
\end{figure}

To achieve the required level of statistical and
systematic uncertainty in $\EPR$, the KTeV experiment 
(Fig.~\ref{fi:ktev}) uses the same double-beam technique as E731 with
a new detector and beamline. Following the primary target, collimators and
sweeping magnets are used to form two almost parallel 
neutral beams.
A fully active regenerator is placed in one of the 
beams 122m from the
production target, at the upstream end of the decay region, to provide a 
source of $\KS$ for the experiment (this beam is
referred to as the
regenerator beam and the other as the vacuum beam). 
The regenerator (along with a movable absorber that attenuates the
beam hitting the regenerator) is moved from one 
beam to the other each minute to eliminate many possible systematic 
errors in normalization and detector response. 
All four $K \To 2\pi$ decays
are detected simultaneously.   
The detector consists of a 
large vacuum decay region instrumented with photon veto counters, a drift
chamber spectrometer, and a CsI electromagnetic calorimeter.
Compared to E731, KTeV also has an improved trigger
and data acquisition system. The final stage of the 
trigger
includes full event reconstruction and filtering before
data are written to tape.

\begin{figure}[ht]	
\centerline{\epsfysize 4 truein 
\epsfbox{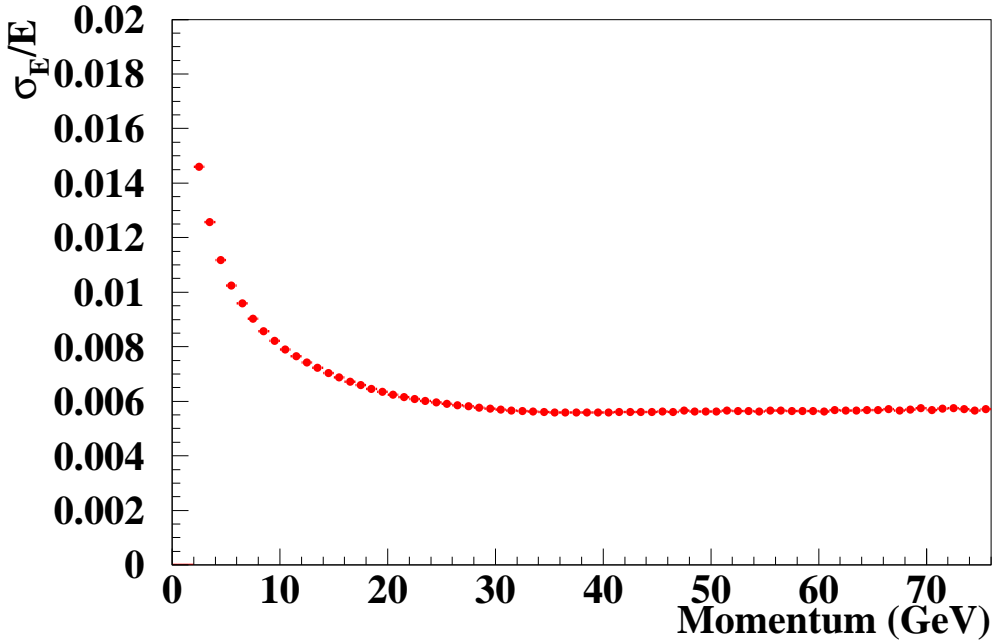}}
\vskip -0.5 cm
\caption[]{
CsI energy resolution as a function of momentum measured with 
200 million electrons
from $K \To \pi e \nu$ decays. 
\label{fi:eres}}
\end{figure}

The performance of the calorimeter, made up of 3100 pure CsI crystals,
is particularly important to the success of the experiment.
The energy scale of the calorimeter is 
directly related to the reconstructed  
decay position along the beam ($z$) direction for
$K \To \pi^0\pi^0$  decays, and is
therefore a critical part of understanding the detector's acceptance for 
neutral events. Figure~\ref{fi:eres} 
shows the energy resolution of the calorimeter
as a function of momentum for electrons from $K \To \pi e \nu$ events.
The average energy resolution for photons from $K \To \pi^0\pi^0$ events is
0.7\%.
The excellent energy
resolution also reduces background for both the $\pi^+\pi^-$ and 
$\pi^0\pi^0$ decay modes.

\section{$K \To \pi^+\pi^-$ and $K \To \pi^0\pi^0$ Analysis}

The analysis presented here is based on $K\To \pi^0\pi^0$ data collected 
during 1996 and $K \To \pi^+\pi^-$ data collected during 
1997~\cite{no96}.
Using charged and neutral data from different running periods does not 
significantly increase the systematic error in $\EPR$
because the two decay modes use
essentially independent detector systems.
It is
critical, however, that the $\KS/\KL$ flux ratio be the same in both years.
This issue
will be addressed in Section 4.

$K \To \pi^+\pi^-$ events are reconstructed in 
the charged spectrometer, consisting of four drift chambers, two on
either side of a large dipole magnet providing a 0.41 GeV/c horizontal momentum
kick. 
Important requirements for the $\pi^+\pi^-$ decay mode include:
\begin{itemize}
\item each track must have a momentum of at least 8 GeV/c and
deposit less than 85\%\ of its energy in the CsI calorimeter;
\item there must not be any 
hits in the muon hodoscope located behind 4 m of steel;
\item the square of the transverse momentum of the $\pi^+\pi^-$ system relative
to the initial kaon trajectory, $\ptsq$, is required to be less than
250 MeV$^2$/c$^2$.
\end{itemize}

\pagebreak

The $K \To \pi^0\pi^0$ reconstruction is based on the energies and positions
of four photons measured in the CsI calorimeter. The events are reconstructed
by 
selecting the photon pairing which is most consistent with both
$\pi^0$ decays occurring at the same point. As an alternative to $\ptsq$, which
cannot be reconstructed in $\pi^0\pi^0$ decays because the 
photon directions are
not measured,
we calculate a ``ring number'' based on the
center-of-energy of the four photons at the calorimeter. The ring number is
defined as the area (in cm$^2$) 
of the smallest square that is centered on the
beam and contains
the center-of-energy. We require that the ring number be less than 110, which
selects events with center-of-energy inside a square region of area 
110 cm$^2$ centered on each beam.

\begin{figure}
\vspace*{-0.3in}
\hspace*{0.05in}
\mbox{\epsfysize=6in \epsffile{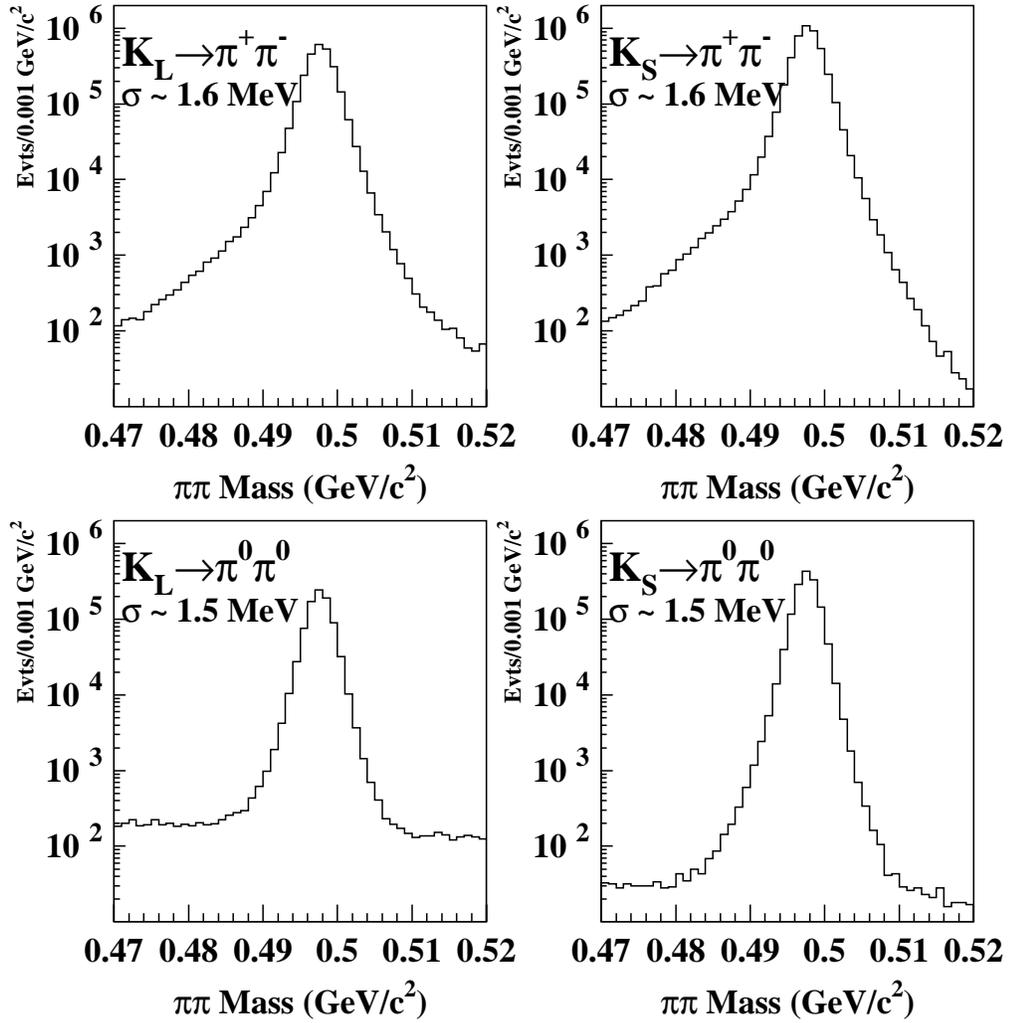}
 } \\
\vspace*{-0.2in}
\caption{$K \To \pi\pi$ invariant-mass plots before background subtraction.
\label{fi:mass}}
\vspace*{0.4cm}
\end{figure}

\begin{figure}	
\centerline{\epsfxsize 5 truein 
\epsfbox{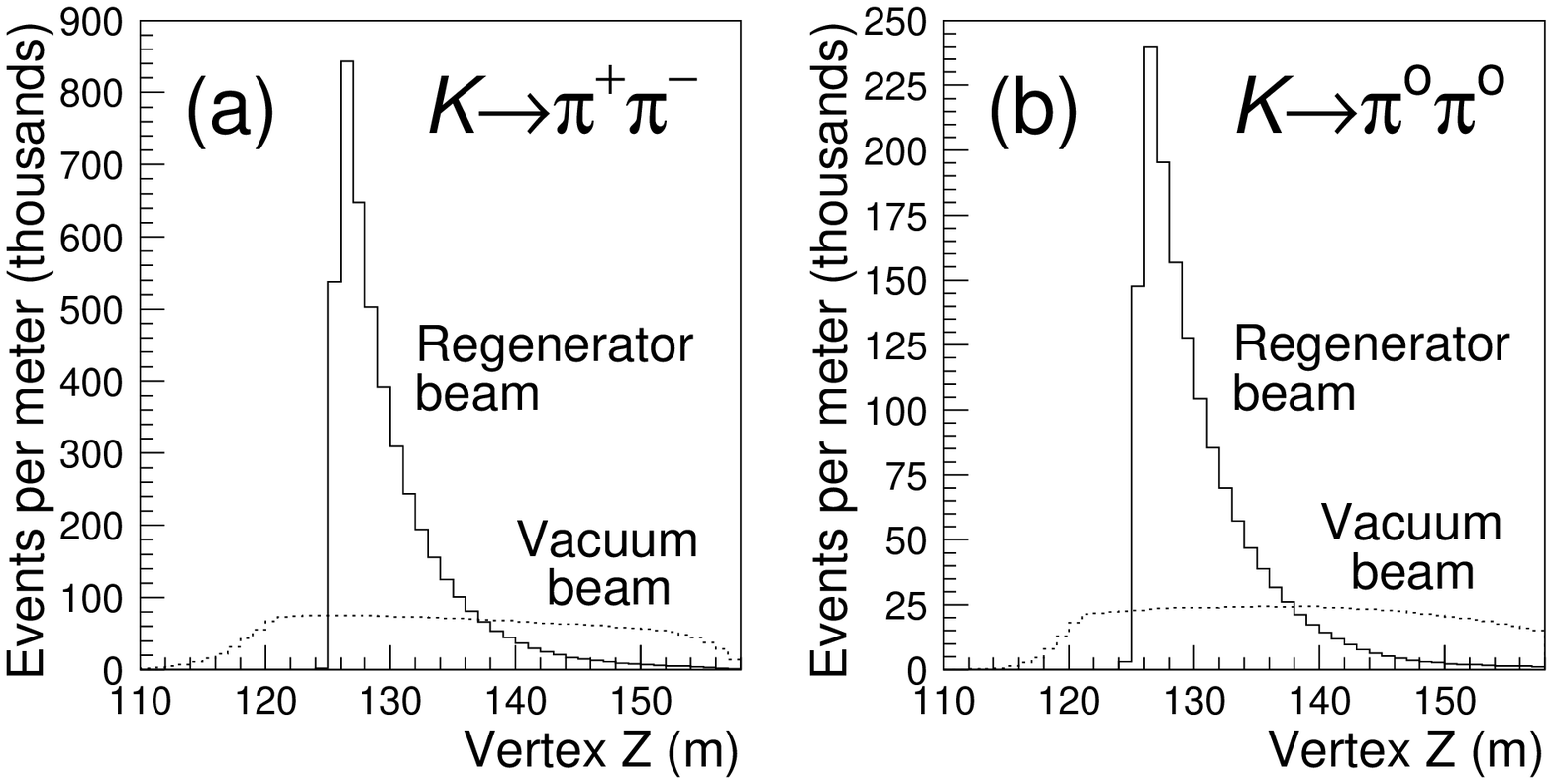}}
\vskip 0.1 cm
\caption[]{
Decay vertex distributions for (a) $K \To \pi^+\pi^-$ and
(b) $K \To \pi^0\pi^0$ decays, showing the difference between the
``regenerator'' ($\KS$) and ``vacuum'' ($\KL$) beams.
\label{fi:zktev}}
\end{figure}

Invariant mass distributions for the $K \To 2\pi$
decay modes for events with $110 < z < 158$ m and $40 < E_K < 160$ GeV are
shown in Fig.~\ref{fi:mass}. 
The corresponding distributions of decay positions along
the beam ($z$) direction are shown in Fig.~\ref{fi:zktev}.

\begin{figure}
\centerline{\epsfxsize 6.0 truein 
\epsfbox{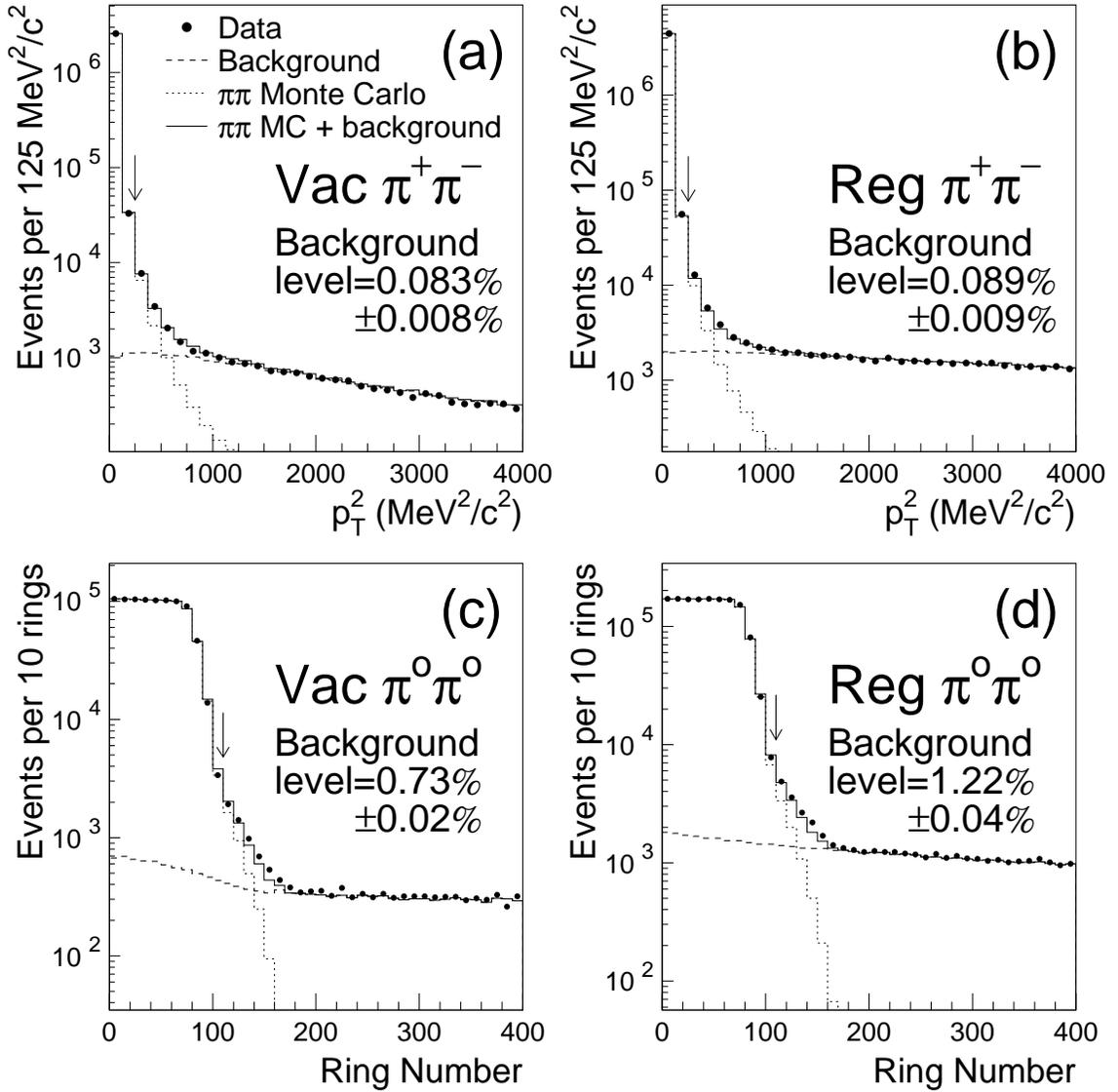}}   
\caption{
Distributions of $\ptsq$ for the $\pi^+\pi^-$ samples and ring number
for the $\pi^0\pi^0$ samples.
Total background levels and uncertainties
are given for the samples passing the analysis requirements indicated
with arrows.
\label{fi:back}}
\end{figure}

There are two classes of background in these $K\To\pi\pi$ samples:
misidentified kaon decays and real $K \To \pi\pi$ events that
have scattered in the regenerator or final collimator.
The total background levels for the four decay modes are shown in
Fig.~\ref{fi:back};  Table~\ref{ta:back} summarizes the 
different components of the background.
For the $\pi^+\pi^-$ decay 
mode, backgrounds in both beams are less than 0.1\%. In
the vacuum beam, the background comes mainly from misidentified semileptonic
decays. In the regenerator beam, the main background is from kaons that 
scatter in the regenerator before decaying to $\pi^+\pi^-$.
The background levels are much higher for the 2$\pi^0$ decay mode. As in the
charged decay mode, kaons that scatter in the regenerator are the main
background in the regenerator beam.  Since the ring-number variable is
not as effective as $p_t^2$ at identifying scattered kaons, however, the 
neutral-mode background from this source is 1.07\%\ 
(more than an order of magnitude
larger than in the charged decay mode).
Kaons that scatter enough to be reconstructed
in the wrong beam contribute a background of 
0.3\%\ to neutral decays in the vacuum beam.
The vacuum beam also has a background of 0.27\%\ from
$\KL\To 3\pi^0$ decays with lost and/or overlapping photons.
The numbers of $K \To 2\pi$ events after background subtraction are given in
Table~\ref{ta:yield}.

\begin{table}[hbt]
\centering
\begin{tabular}{|l|cccc| }  \hline  
 & Vac. Beam  & Reg. Beam& Vac. Beam 
& Reg. Beam \\ 
 & $\KL \To \pi^+\pi^-$ & $\KS \To \pi^+\pi^-$ & $\KL \To \pi^0\pi^0$ & $\KS \To \pi^0\pi^0$ \\ \hline
Misidentified K: & & & & \\
\hspace*{0.25in} $K \To \pi\ell\nu$ & 0.069  & 0.003 & & \\
\hspace*{0.25in} $K \To 3\pi^0$ & & & 0.27 & 0.01 \\
Scattered K: & & & & \\
\hspace*{0.25in} regenerator & & 0.072 & 0.30 & 1.07 \\
\hspace*{0.25in} collimator & 0.014 & 0.014 & 0.16 & 0.14 \\
\hline
Total & 0.083 & 0.089 & 0.73 & 1.22 \\ \hline
\end{tabular}
\caption{Background levels (in \%) for $K \To \pi^+\pi^-$ and $K \To \pi^0\pi^0$}
\label{ta:back}
\end{table}

\begin{table}[hbt]
\centering
\begin{tabular}{|l|c|c| }  \hline  
 & Vacuum Beam ($\KL$) & Regenerator Beam ($\KS$) \\ \hline
$K \To \pi^+\pi^-$ & 2,607,274 & 4,515,928 \\
$K \To \pi^0\pi^0$ & 862,254 & 1,433,923 \\
\hline
\end{tabular}
\caption{Numbers of events after background subtraction}
\label{ta:yield}
\end{table}

As shown in Fig.~\ref{fi:zktev}, 
the difference between the $\KL$ and $\KS$
lifetimes results in very different decay vertex distributions for the
$\KL$ and $\KS$ decays which must be compared to compute $\EPR$.
Therefore,
to extract a value for $\EPR$, the numbers in Table~\ref{ta:yield} must be
corrected for the variation in detector acceptance as a function of $z$. 
We make this correction with a Monte Carlo (MC) simulation. 
The simulation models kaon production
and regeneration to generate decays with the same energy and $z$ distribution
as the data.  It also includes a detailed simulation of all detector elements.

\begin{figure}	
\vspace*{1in}
\centerline{\epsfxsize 6.0 truein 
\epsfbox{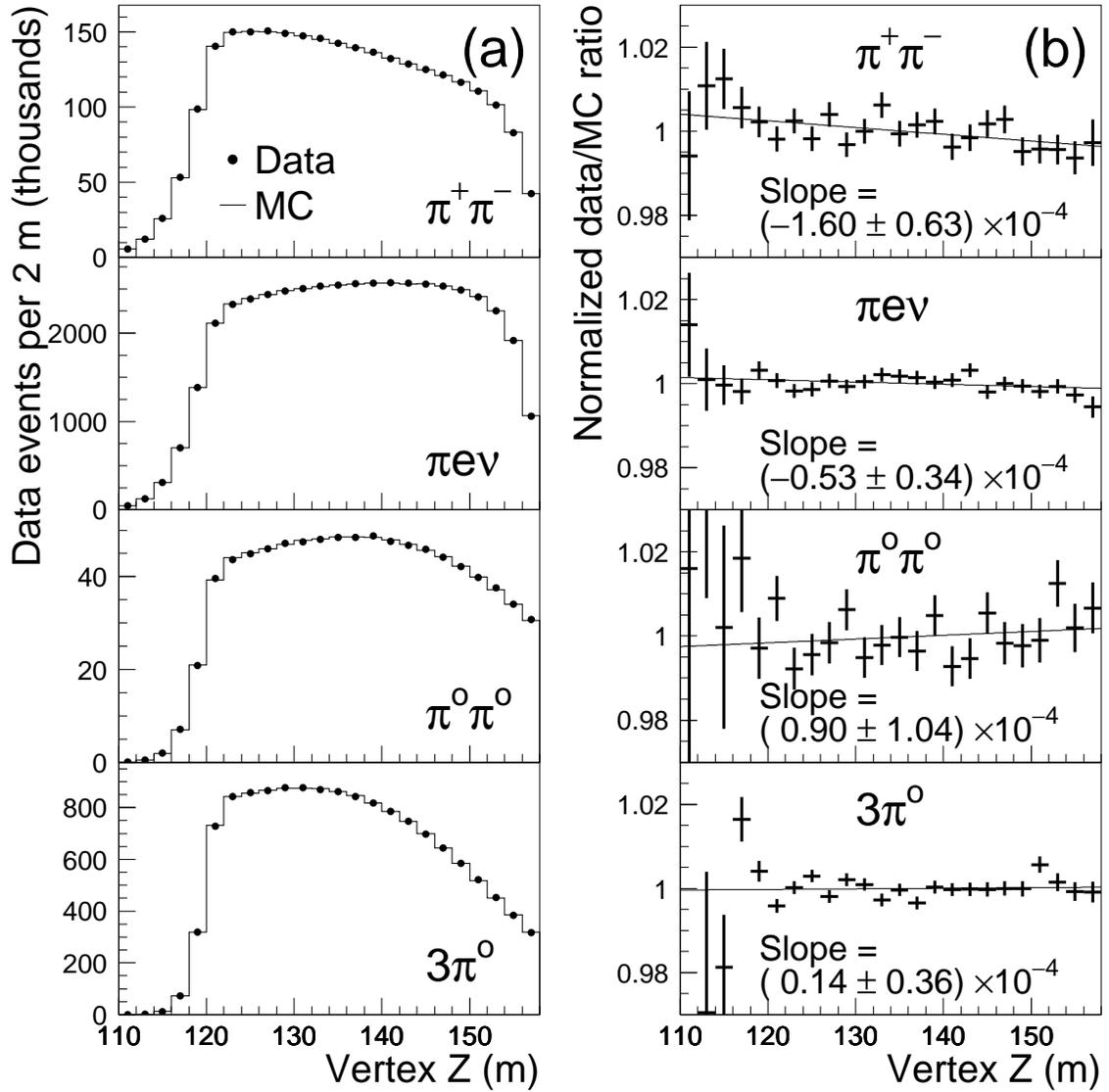}}   
\vskip 0.5 cm
\caption[]{(a) Data -- Monte Carlo comparisons of vacuum-beam $z$ 
distributions for $\pi^+\pi^-$, $\pi e\nu$, $\pi^0\pi^0$, and $3\pi^0$
decays; (b) straight-line fits to the data/MC ratios.
\label{fi:zslopes}
}
\end{figure}

The
quality of the Monte Carlo 
simulation is studied using distributions from $K \To 
2\pi$
decays, as well as higher statistics decay modes.
Figure~\ref{fi:zslopes} 
shows a comparison of data and Monte Carlo vacuum beam $z$ vertex
distributions for the $\pi\pi$ signal modes, as well as for the much larger
$\pi e \nu$ and
$3\pi^0$ samples. Since the average decay positions for $\KL$ and $\KS$ decays
differ by about 6 m, 
a relative slope of $10^{-4}$ per meter in the data/MC ratio
would result in an error of $10^{-4}$ in $\EPR$. 
The only noticeable problem in Fig.~\ref{fi:zslopes}
is the slope of $(-1.60 \pm 0.63) \times 10^{-4}$ m$^{-1}$ 
in the data/MC slope for $K \To \pi^+\pi^-$. Although this slope is 
only 2.5 sigma from zero and the
slope in $K_L \To \pi e \nu$ is much smaller, we assign
a systematic error of $1.6 \times 10^{-4}$ on Re($\EPR$)
based on the full size of the observed slope in 
$\KL \To \pi^+\pi^-$. The $\pi^0\pi^0$ and $3\pi^0$ data and Monte Carlo
$z$ distributions are consistent.  
Since the $3\pi^0$ decay mode is more sensitive
to most acceptance problems, we use the slope in the data/MC 
ratio in $3\pi^0$ to place a limit of $0.7 \pm 10^{-4}$ on the bias in
Re$(\EPR)$ from the neutral-mode acceptance.

\section{Systematic Errors}

Table~\ref{ta:syst} summarizes the estimated systematic uncertainties
in $\EPR$.  The $\pi^+\pi^-$ and $\pi^0\pi^0$ columns 
list errors that are specific to the charged
and neutral decay modes.  The errors at the bottom of the table are common
to both decay modes.  Adding these errors in quadrature yields a total
systematic error on $\EPR$ of $2.8 \times 10^{-4}$. Only a few of these errors
will be addressed here; additional details are given 
in~\cite{Alavi-Harati:1999xp}.

The largest contribution to the systematic error comes from
uncertainties in the $z$ dependence of the acceptance, which
are estimated from the comparison of
$z$ distributions for data and Monte Carlo discussed above. 
Other significant
uncertainties result from the energy scale and background subtraction
in neutral mode.  
To evaluate the uncertainty in the neutral energy scale,
we compare the reconstructed $z$ vertex of 
$\pi^0\pi^0$ events 
produced at the downstream edge of the regenerator with the 
reconstructed $z$ vertex for $\pi^0$ pairs produced by hadronic interactions
in the vacuum window at the downstream end of the decay region. The 
difference between these measurements 
is 2 cm greater than the actual distance
between the regenerator edge and the vacuum window, resulting in a systematic
error of $0.7 \times 10^{-4}$ on Re($\EPR$).
The neutral mode background uncertainty results largely from uncertainty in 
the acceptance for scattered
$K \To \pi^+\pi^-$ decays, which are used to tune
the Monte Carlo simulation for scattered events.

As mentioned earlier, in combining $\pi^+\pi^-$ and
$\pi^0\pi^0$ data from different years, we must consider the possibility
of a change in the $\KS/\KL$ flux ratio between the two samples. Although
the same
regenerator and movable absorber were used for the two data samples, 
we assign a small uncertainty on Re$(\EPR)$ to account for
the possible effect 
of a temperature difference between the two data collection periods, 
which would change
the densities of the movable absorber and regenerator.

\begin{table}
\centering
  \begin{tabular}{|l|c|c|} \hline
            & \multicolumn{2}{c|}{Uncertainty ($\times 10^{-4})$} \\
 Source of uncertainty &  from $\pi^+\pi^-$  &  from $\pi^0\pi^0$  \\
\hline 
\hline 
Trigger (L1/L2/L3)   & 0.5 & 0.3 \\
Energy scale & 0.1 & 0.7 \\
Calorimeter nonlinearity & --- & 0.6 \\
Detector calibration, alignment &  0.3 & 0.4 \\
Analysis cut variations & 0.6 & 0.8 \\
Background subtraction & 0.2 & 0.8 \\
Limiting apertures  & 0.3 & 0.5 \\   
Detector resolution  &  0.4 & 0.1 \\
Drift chamber simulation & 0.6 & --- \\
$Z$ dependence of acceptance &  1.6  &  0.7 \\
Monte Carlo statistics & 0.5 &  0.9  \\ \cline{2-3}
{$\KS / \KL$ flux ratio:} & \multicolumn{2}{|c|}{ } \\
   \hspace*{0.25in} 1996 versus 1997 & \multicolumn{2}{c|}{ $0.2$} \\
   \hspace*{0.25in} Energy dependence & \multicolumn{2}{c|}{ $0.2$} \\
$\Delta m$, $\tau_S$, regeneration phase  &
                                       \multicolumn{2}{c|}{ $0.2$} \\
\hline\hline 
 TOTAL  &                              \multicolumn{2}{c|}{ $2.8$} \\ \hline
\end{tabular} 

\caption{Summary of systematic uncertainties in Re($\EPR$)
        }
\label{ta:syst}
\end{table}

\section{Results}

\begin{figure}	
\centerline{\epsfxsize 4.5 truein 
\epsfbox{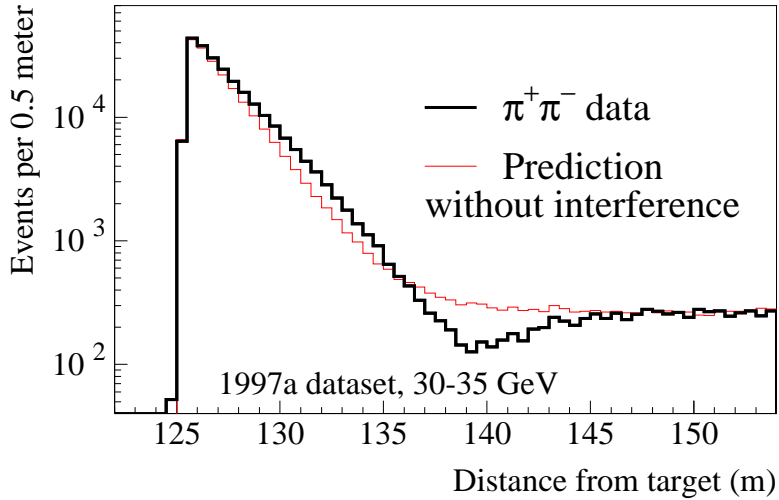}}   
\vskip 0 cm
\caption[]{Decay vertex distribution from $K \To \pi^+\pi^-$ events with
$30 < E_K < 35$ GeV downstream of the regenerator showing $\KS-\KL$
interference.
\label{fi:interference}
}
\end{figure}

The numbers of events and relative acceptances (from the Monte Carlo 
simulation) for the four 2$\pi$ decay modes can be used to calculate a
simple estimate of $\EPR$ using Equation 1.  This calculation~\cite{fineprint} 
yields ${\rm Re}(\EPR) = (26.5 \pm 3.0) \times 10^{-4},$
but is not precisely 
correct because it assumes that there are only $\KS$ decays
in the regenerator beam.  As illustrated in Fig.~\ref{fi:interference}, the
regenerator beam contains 
a coherent mixture of $\KS$ and $\KL$, which
must be taken into account in the calculation of $\EPR$.  The decay 
distibution in the regenerator beam also allows us to measure the
$\KS$ lifetime ($\tau_S$), 
the $\KS-\KL$ mass difference ($\Delta m$), and the relative phases
of the CP violating and CP conserving amplitudes ($\Phi_{00}$ for
$K\To\pi^0\pi^0$ and $\Phi_{++}$ for $K\To\pi^+\pi^-$):
\begin{eqnarray*}
\tau_S & = & (0.8967 \pm 0.0007) \times 10^{-10} s \\
\Delta m & = & (0.5280 \pm 0.0013) \times 10^{-10} \hbar s^{-1}\\
\Delta \Phi & = & \Phi_{00} - \Phi_{+-} = 0.09^\circ \pm 0.46^\circ .
\end{eqnarray*}

The final value of Re$(\EPR)$ is extracted from the background-subtracted 
and \linebreak 
acceptance-corrected data with a fitting 
program that calculates decay vertex distributions, properly treating 
regeneration and $\KS-\KL$ interference downstream of the regenerator.
Including the systematic error from Table~\ref{ta:syst}, 
the result of the 
fit is
\begin{eqnarray*}
{\rm Re}(\EPR) & = &(28.0 \pm 3.0\ ({\rm stat}) 
\pm 2.8\ ({\rm syst})) \times 10^{-4} \\
 & = & (28.0 \pm 4.1 ) \times 10^{-4}.
\end{eqnarray*}

\begin{figure}	
\centerline{\epsfxsize 4.2 truein 
\epsfbox{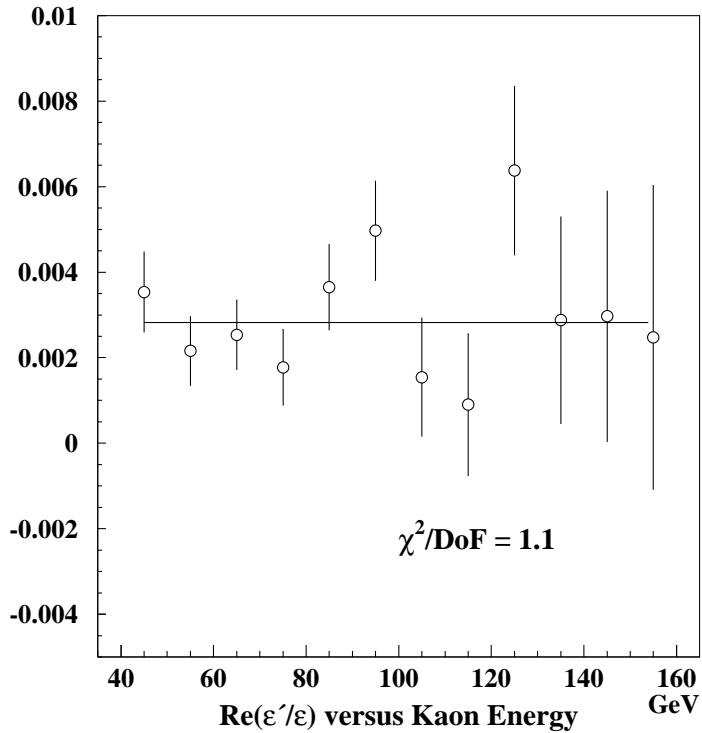}}   
\vskip 0 cm
\caption[]{
Re$(\EPR)$ as a function of kaon energy. The horizontal line represents
the central value of Re$(\EPR)=28.0 \times 10^{-4}$. Only statistical errors
are shown.\label{fi:pk}
}
\end{figure}

\begin{figure}	
\centerline{\epsfxsize 3.5 truein 
\epsfbox{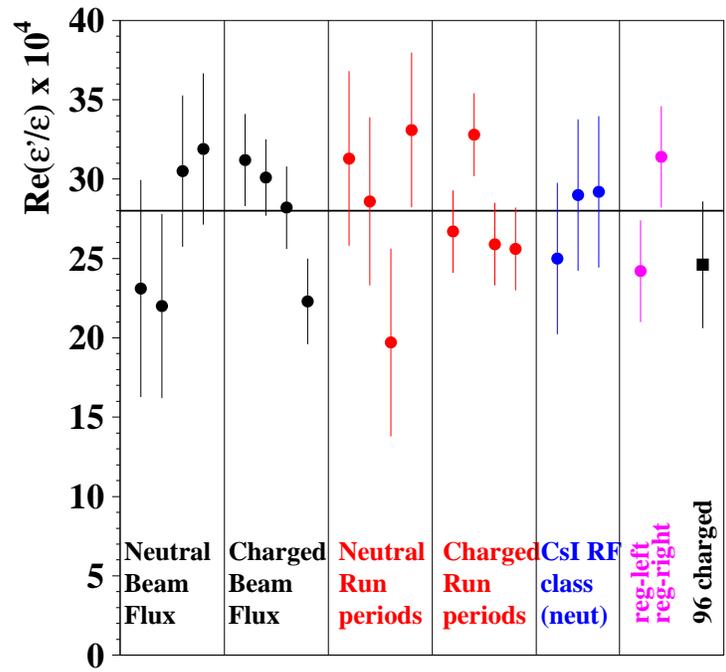}}   
\vskip -2.0 cm
\caption[]{
Cross checks on $\EPR$ measurement. The horizontal line represents
the central value of Re$(\EPR)=28.0 \times 10^{-4}$. Within each category,
only independent statistical errors are shown.  The ``96 charged'' point
also includes the additional systematic error described in~\cite{no96}.
\label{fi:checks}
}
\end{figure}

We have performed several cross checks on this result. Consistent results
are obtained at all kaon energies, beam intensities, periods during the run,
magnet polarities, and for both regenerator positions. We also have done
the analysis using $\pi^+\pi^-$ data from 1996 (collected simultaneously with
the $\pi^0\pi^0$ data) rather than $\pi^+\pi^-$ data from 1997. The result
is consistent with that quoted above, but with a larger systematic 
error~\cite{no96}. Some of these cross checks are summarized in 
Figs.~\ref{fi:pk} and~\ref{fi:checks}.  It is worth noting that the $\EPR$
analysis was done ``blind'': the value of $\EPR$ was hidden with an unknown
offset until after the analysis and evaluation of the systematic error were
completed.

\section{Conclusions}

\begin{figure}	
\centerline{\epsfxsize 4.6 truein 
\epsfbox{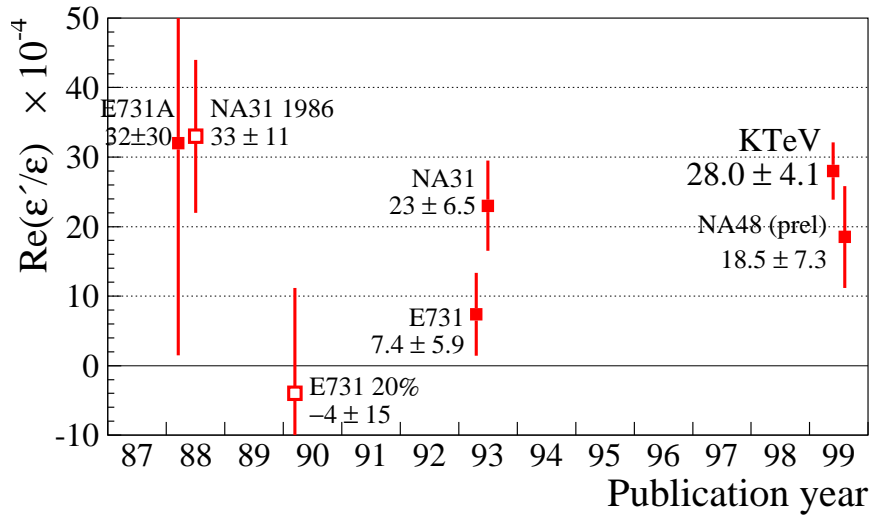}}
\vskip 0.0 cm
\caption[]{
\label{fi:year}
Recent measurements of $\EPR$. The 
solid points are statistically independent; the
open points are included in other measurements.}
\end{figure}

Based on about 1/4 
of the data collected during the Fermilab 1996-1997 fixed-target
run, KTeV has measured Re$(\EPR) = (28.0 \pm 3.0\ ({\rm stat}) 
\pm 2.8\ ({\rm syst})) \times 10^{-4}$.
This result establishes the existence of CP violation in a decay process at
almost 7 sigma, and rules out the Superweak Model as the sole source of CP
violation. 
Although the result is larger than most Standard Model calculations,
it supports the Standard Model explanation of CP violation.
Figure~\ref{fi:year} 
shows a comparison of the KTeV measurement with previous 
results; the preliminary
result of the NA48 experiment~\cite{barr,na48pub} is shown also.
The KTeV result
is consistent with earlier evidence for direct CP 
violation from NA31 and differs from the E731 result by 2.9 sigma. Study
of the E731 result has not revealed any error that would cause
this discrepancy.  A weighted average of all $\EPR$ measurements gives
Re$(\EPR) = (21.3 \pm 2.8) \times 10^{-4}$ with a confidence level of
7\%.

The analysis of the remaining 3/4 of KTeV's 1996-1997 data sample is in 
progress.  In 1999, we collected a new $\EPR$ data set with statistics
equal to the full 1996-1997 sample.  Several small detector modifications
were made to improve the systematic quality of the 1999 data.  We also
performed additional systematic studies during the 1999 run.
The full data sample should allow
us to reduce the statistical uncertainty on 
Re$(\EPR)$ to $1\times 10^{-4}$. Significant work will be required to reduce 
the systematic error to a similar level.

Although the very 
existence of direct CP violation has been an issue until recently,
the full data sets of KTeV, NA48, and KLOE may soon provide measurements
of $\EPR$ at the 5\% level.  Considerable
improvement in theoretical calculations of $\EPR$ will be required to take 
full advantage of this precision. There is some optimism, however,
that the next rounds of
calculations using lattice gauge theory may approach a 10\% uncertainty
in $\EPR$, making the precise measurements of $\EPR$
equally precise tests of
the Standard Model.

\newpage

\def\Discussion{
\setlength{\parskip}{0.3cm}\setlength{\parindent}{0.0cm}
     \bigskip\bigskip      {\Large {\bf Discussion}} \bigskip}
\def\speaker#1{{\bf #1:}\ }

\Discussion

\speaker{Sherwood Parker (University of Hawaii)}
Did you make any changes to the apparatus to achieve the better results
in the second run?

\speaker{Blucher}  There were no fundamental changes to the detector, but
small modifications were made to improve data quality and datataking 
efficiency. For data quality, the most important changes involved the
drift chamber system.
The drift chamber electronics
were improved to have higher gain and lower
noise to reduce the chamber inefficiency. The two upstream drift chambers
were also restrung because of damage sustained during the 1996-1997 run.

\speaker{B.~F.~L. Ward (University of Tennessee)}
Is there any understanding of why there may have been a systematic
bias in the earlier
measurement that caused it to miss the higher value of
Re$(\epsilon^\prime/\epsilon)$?

\speaker{Blucher}
As mentioned earler, 
we have studied the E731 analysis and have not found any evidence 
of a systematic problem. For example,
comparisons of data and Monte Carlo $z$ distributions for 
E731~\cite{Gibbons:1997fw}, 
similar to
those shown for KTeV in Fig.~\ref{fi:zslopes}, do not show any sign of an
acceptance problem.

\speaker{Mario Calvetti (INFN, Florence)}
In your analysis, you  rely on  the Monte Carlo
 acceptance corrections to the first order.  Why is this
preferable to the method of NA48; that is,
reweighting the events in order
to have similar longitudinal vertex distribution?

\speaker{Blucher}
The NA48 procedure sacrifices statistics to reduce the required acceptance
correction, while
the KTeV procedure maximizes the statistical power of the data sample, but 
requires that the detector acceptance be understood.
We believe that we have a reliable procedure to estimate the systematic
error
associated with our acceptance correction.  As a cross check,
we have analyzed our data
with an alternate technique that compares vacuum and
regenerator beam $z$ distributions
directly, eliminating the need for a
Monte Carlo  acceptance correction.  This analysis gives a
consistent value of $\epsilon^\prime/\epsilon$ but with a
significantly larger statistical error.


\begin{thebibliography}{99}
\bibitem{fineprint}For this estimate, the $z$ range for
decays in the regenerator beam
is restricted to the region just downstream of the regenerator 
($124 < z < 132$ m) to reduce the importance of $\KL$ decays in that beam.
\end{thebibliography}
\end{document}